\documentclass{article}
\usepackage{graphicx} 
\usepackage{geometry}
\usepackage{amsmath}
\usepackage[utf8]{inputenc}
\usepackage{amssymb}
\usepackage{amsfonts}
\usepackage{amsthm}
\usepackage{mathrsfs}
\usepackage{color}
\usepackage{comment}
\usepackage{algpseudocode}
\usepackage{algorithm}
\usepackage{placeins}
\usepackage{appendix}
\usepackage{fullpage}

\DeclareMathOperator*{\argmin}{arg \, min}

\newcommand{\R}{\mathbb{R}}
\newcommand{\X}{\mathbf{X}}
\newcommand{\rr}{\mathbf{R}}

\newcommand{\Mm}{K}

\newcommand{\bbO}{\mathbb{O}}
\newcommand{\bx}{\mathbf{x}}
\newcommand{\br}{\mathbf{r}}

\newcommand{\xt}{{\tilde{x}}}
\newcommand{\Nt}{{\tilde{N}}}

\newcommand\itc[1]{#1}

\newcommand{\rev}[1]{%
  \begingroup
  #1%
  \endgroup
}

\title{Using Echo-State Networks to Reproduce Rare Events in Chaotic Systems}
\author{Anton Erofeev\thanks{\texttt{University of Houston, aerofeev@uh.edu}}, 
Balasubramanya T. Nadiga\thanks{\texttt{Los Alamos National Laboratory, balu@lanl.gov}}, 
Ilya Timofeyev\thanks{\texttt{University of Houston, itimofey@Central.uh.edu}}}
\date{July 2026}

\begin{document}

\maketitle

\begin{abstract}
\rev{
Echo-State Networks (ESNs) are known to reproduce trajectories and main properties of chaotic attractors. Here, we investigate the question of whether ESNs can reproduce rare events in a system with a stationary distribution with slowly decaying bounded tails. In particular, we consider the chaotic four-dimensional competitive Lotka--Volterra system, and quantify extremes using the Generalized Extreme Value distribution. In stationary simulations, the ESN reproduces the stationary distributions and rare event statistics with high accuracy. 
In addition, the ESN also reproduces rare event statistics for non-equilibrium ensemble simulations for initial conditions relatively close to the underlying attractor.
However, the performance of the ESN deteriorates as the ensemble spread increases and trajectories explore regions of phase space not represented in the training data.}  
\end{abstract}


\section{Introduction}
Machine learning has emerged as an alternative approach for solving partial differential equations, reproducing trajectories of dynamical systems, emulating statistical properties of chaotic systems, etc. Neural networks and deep learning play a particularly important role in developing new techniques for understanding and solving various dynamical systems. 

Reservoir computing \cite{jaeger2004harnessing, esp1}
is a particular class of machine learning models; it utilizes a large recurrent network 
(reservoir), and only a linear output layer is trained to match the trajectory. 
Echo-State Networks refer to reservoirs that have the Echo-State Property
(see e.g. \cite{jaeger2001echo,jaeger2002NIPS,maass2002real,buehner2006,jaeger2007echo,esp3, esp1}).
There has been a considerable amount of research on Reservoir Computing (RC) and Echo-State Networks (ESNs). In particular, 
it has been demonstrated that ESNs are able to predict time series of nonlinear and chaotic models for several Lyapunov times (e.g. \cite{pathak2017using, estevez2023echo, hesch21, girvan2020separation, nadiga2021reservoir, chen2022predicting, doan2020physics, gonzalez2022optimizing,platt2022}).
Moreover, starting with \cite{pathak2017using}, 
it has also been demonstrated that Echo-State Networks are capable of reproducing the statistical properties of attractors in 
chaotic systems (see e.g. \cite{pathak2018model, zhixin2018attractor,vlachas2020attractor,LI2022321,antonik2018using,racca2021robust,tanaka2019recent,barbosa2022chaos}).
\rev{Thus, reproducing an invariant measure or bulk attractor statistics is not, by itself, a new capability of ESNs.}
\rev{Agreement in bulk statistics, however, does not necessarily imply accurate reproduction of distributional tails or rare events, which are sparsely represented in the training data and may therefore receive little weight during training. This distinction is particularly important in geophysical, ecological, and other applications in which extreme events may play an important role. 
These applications motivate the present analysis, but we restrict our investigation to the four-dimensional chaotic Lotka--Volterra system and do not investigate the scalability of ESNs to high-dimensional models.
Therefore, our goal is to demonstrate that ESNs can reproduce tail-sensitive statistics for a chaotic system whose distributions differ from the Lorenz--63 and Kuramoto--Sivashinsky examples. Specifically, we consider the competitive Lotka--Volterra system, whose marginal stationary distributions exhibit bounded but slowly decaying tails, and evaluate both stationary extreme-value statistics and finite-time extremes in transient ensemble simulations, paying particular attention to the conditions under which the ESN performs well, or its performance deteriorates.}

We use Echo-State Networks to predict the dynamics of the four-dimensional competitive Lotka-Volterra (LV) equations \cite{book:elegantchaos, Vano2006lv4, wang2010lv4} in the chaotic regime. 
\rev{We assume that the LV system considered here admits an attractor with an ergodic invariant measure $\mu$, so that the long-time statistical behavior of its trajectories is characterized by this measure. In the chaotic regime considered here, the marginal distributions associated with $\mu$ have sharp peaks and bounded but slowly decaying tails, unlike those of the Lorenz--63 and Kuramoto--Sivashinsky examples.
Since the metric for the ESN's training focuses on individual trajectories, it is not obvious that the ESN can learn the tails of the invariant measure $\mu$ and corresponding statistics of rare events. 
We use Generalized Extreme Value theory to compare statistics of extremes in both the equilibrium and non-equilibrium settings.
For the stationary regime, 
we demonstrate that the ESN learn the tails of the invariant measure and reproduce statistics of rare events with high accuracy.
For the non-equilibrium regime, the performance of the ESN deteriorates as the ensemble of initial conditions becomes more spread out, and 
trajectories visit regions of phase space away from the underlying attractor.}
Reproducing rare events is a notoriously difficult problem since the amount of training data is much smaller for these regions of phase space. Therefore, many
Machine Learning and other types of reduced or surrogate models tend to reproduce bulk (e.g., means, variances, etc.) statistical properties correctly. 
\rev{However, it is not obvious that Machine Learning models are capable of accurately capturing rare events.
Studying this question in the context of ESNs is the main novel contribution of this paper.}

\rev{The paper is organized as follows.}
First, we carry out a standard test and demonstrate that ESNs can predict individual trajectories of the 4D LV system for several Lyapunov times.
We also discuss the consequences of generalized synchronization for the ESN trained on the LV data.
Then, we carry out a comparison of the equilibrium statistical simulations and quantitatively demonstrate that the ESN reproduces the complex structure of the invariant measure. \rev{Moreover, we define rare events as maxima over non-overlapping blocks, and demonstrate that the ESN 
reproduces the statistics of rare events and tails of marginal stationary distributions with high accuracy.} We use the generalized extreme value (GEV) distribution to quantify the performance of the ESN with respect to rare events. 
We also demonstrate numerically that the ESN reproduces rare events in ensemble simulations of the LV model. However, it is important to note that the ESN's performance is adequate when the initial ensemble is centered on the attractor of the LV model. For ensembles centered away from the attractor, the ESN's performance deteriorates since the model is not trained on solutions in that part of the phase space.
\rev{We also find that the performance of the ESN deteriorates as the initial variance of the ensemble increases, even when the ensemble is centered on the attractor.}
Therefore, we can conjecture that there is an upper bound on the ESNs' predictive power for ensemble simulations.
\rev{We also analyze the performance of the ESN with incomplete observations and compare the performance of the ESN with the Next Generation Reservoir Computing (NGRC) \cite{gauthier2021ngrc}.}

\section{Dynamic Equations and the Reservoir}
Competitive Lotka-Volterra (LV) equations are a model 
of population dynamics for species competing for the same resource. It is a simple generalization of the classical 2D predator-prey model.
We consider the 4D competitive Lotka-Volterra system
\begin{equation}
\label{lv}
\frac{d}{dt} x_i = r_i x_i \left( 1 - \sum\limits_{j=0}^3 \alpha_{ij} x_j \right), 
\quad i=0,\ldots,3
\end{equation}
with
\[
r = \left( 
\begin{array}{c}
1 \\ 0.72 \\ 1.53 \\ 1.27
\end{array}
\right), \qquad
\alpha = \left( 
\begin{array}{cccc}
1 & 1.09 & 1.52 & 0 \\
0 & 1    & 0.44 & 1.36 \\
2.33 & 0 & 1 & 0.47 \\
1.21 & 0.51 & 0.35 & 1
\end{array}
\right).
\]
The above model was investigated in \cite{Vano2006lv4,wang2010lv4}.
It was demonstrated that this model exhibits chaotic behavior 
for this set of parameters. Moreover, the chaotic behavior is robust to perturbations but occurs for a relatively narrow set of parameters. The largest Lyapunov exponent for this model is $\lambda_{max} =  0.0203$. Thus, the Lyapunov time is approximately $T_{lyap} \approx 49.26$.
In this paper,
this model is integrated numerically using a fixed time-step Runge-Kutta 4th-order method with the time-step 
$\delta t = 0.01$. We sample trajectories with the time-step $\Delta t=2$. Thus, $\Delta t=2$ is used as the ESN time-step for training and generating trajectories.

\textbf{Reservoir.}
We use standard reservoir architecture. The equations for the reservoir are given by 
\begin{equation} \label{res}
  r(t+\Delta t) = f(A \, r(t) + W_{in} \, x(t)),  
\end{equation}
where $r(t) \in \R^D$ is the reservoir state, $f = \tanh(x):\R^D \to \R^D$ is the element-wise activation function.
Matrices $A \in \R^{D \times D}$ and $W_{in} \in \R^{D \times N}$ ($N=4$ is the dimension of the LV system) remain constant throughout the training process.
During training, only the weights of the output layer, $W_{out} \in \R^{N \times D}$, are adjusted as linear regression weights for the reservoir states $r(t + \Delta t)$ to produce the output 
$x(t + \Delta t)$. The explicit solution for $W_{out}$ is given by 
\begin{equation}
 W_{out} = \X  \rr' \left( \rr\rr' +\mu I \right)^{-1},
 \label{W_out formula}
\end{equation}
where $\mathbf{X} \in \R^{N \times M}$ is the training data (trajectories) and $\mathbf{R} \in \R^{D \times M}$ are the time-series 
 of the reservoir (with fixed $A$ and $W_{in}$),
 and $\rr'$ denotes the transpose of $\rr$. The size of the training data, $M$ is the number of snapshots generated from simulations of the LV model.
Thus, the prediction is given by
\begin{equation}
\label{ESN}
  \hat{x}\left( t +\Delta t\right) = W_{out} \, r(t+\Delta t).
\end{equation}
During the prediction phase, $\hat{x}(t)$ is used in equation \eqref{res} instead of $x(t)$ to generate the reservoir update; thus, the reservoir becomes an autonomous system.
We do not use a non-linear transformation of the
reservoir in this paper.

\textbf{Reservoir parameters.} We use the reservoir with the size $D=200$ and sparsity $s = 90\%$. We also verified that reservoirs with sparsity $45\%$ and $20\%$ lead to similar results. The spectral radius is 
$\lambda_{res} = 0.5$. The input layer is generated from a uniform distribution $W_{in} \sim Unif[-w,w]$ with 
$w = 5$. The regularization parameter is
$\mu=0.001$. 
The size of the training data
500,000 is sufficient for the accurate reproduction of trajectories and extremes.
For the prediction phase, we warm up the reservoir for 10 steps $\Delta t$.

\section{Numerical Results}
In this section, we present several numerical results 
comparing the behavior of the ESN with the simulations of the LV model. 

\subsection*{Generalized Synchronization}
An interesting aspect of reservoir computing as described above is
that the reservoir is never trained. Furthermore, RC has proven to be
as skillful or more skillful than a number of other more sophisticated
machine learning techniques.

Let
$$
\frac{d\mathbf x(t)}{dt} = \mathbf F_x \left( \mathbf x(t)\right)
$$
represent the dynamical system underlying the data; here $\mathbf F_x
\left( \mathbf x(t)\right)$ can be assumed unknown. Next,
since the reservoir is forced by input $\mathbf x(t)$, the
dynamics of the reservoir are non-autonomous and may be represented as
$$
\frac{d\mathbf r(t)}{dt} = \mathbf F_r \left( \mathbf r(t),
  \mathbf x(t); \boldsymbol\theta \right).
  $$

Generalized synchronization (GS) is said to occur between these two dynamical
systems when they behave in such a way that there is a functional relationship
between their states. That is GS occurs if asymptotically (meaning
after finite transients), $\mathbf r(t)
= \psi(\mathbf x(t))$, where $\psi$ is some unknown nonlinear transformation. In other words, the state of the reservoir
becomes a deterministic function of the driver $\mathbf x(t)$. Unlike
identical synchronization, GS does not require state equivalence,
allowing complex interdependencies.

In \cite{platt2021gensync} it was 
proposed that predictive GS
occurs when $\mathbf x = \phi(\mathbf r)$ without assuming $\phi =
\psi^{-1}$ globally. As such, with the predictive GS,
the non-autonomous dynamics of the reservoir above can be approximated
in an autonomous fashion as
$$
\frac{d\mathbf r(t)}{dt} = \mathbf F_r \left( \mathbf r(t),
\tilde{\boldsymbol\phi}(\mathbf r(t)); \boldsymbol\theta \right).
$$

The auxiliary test in the present context amounts to creating two
copies of a reservoir (so that their connectivity is the same) and
subjecting the two reservoirs to the same input (driver dynamical
system). The only difference between the reservoir systems is that
their initial states are slightly different: $\mathbf r_2(0) = \mathbf
r_1(0) + \epsilon N(0,1)$. GS is validated if
$\mathbf r_2(t) - \mathbf r_1(t)$
decays asymptotically to 0. This amounts to saying that the
predictions must be independent of the initial conditions of the reservoir.

\begin{figure}
\centerline{
\includegraphics[width=0.65\textwidth]{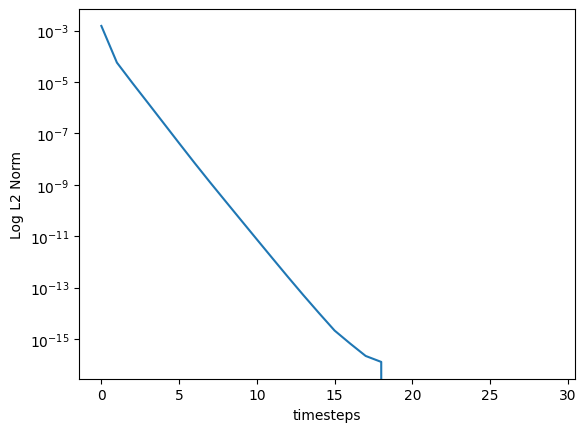}}
  \caption{Log of the $L^2$ norm 
  $||r_1(t) - r_2(t)||_2$ between two reservoirs with different initial conditions. One time-step is $\Delta t=2$. We observe a fast exponential decay of the $L^2$ norm, indicating that the initial conditions are not affecting the utility of predictions for individual trajectories.}
  \label{fig1}
\end{figure}

\textbf{Testing the Independence from Initial Conditions.}
Since we warm up the reservoir, we test whether the reservoir becomes independent of initial conditions during the warm-up phase. To this end, after $W_{in}$ and $A$ are generated, we consider two reservoirs with different initial conditions.
We use the initial condition $r_1(0)=0$ for the first reservoir. The second reservoir is initialized using a normal distribution 
$r_2(0) = \epsilon \times N(0,1)$. Then, the two reservoirs receive the identical input $x(t)$ (generated by the LV model) and should synchronize. Algorithm \ref{alg1} in Appendix \ref{sec:alg} describes this test. We use $\epsilon=0.01$ and confirm that the two reservoirs synchronize for several $W_{in}$
and $A$, including sparsity 45\% and 20\% for the connectivity matrix $A$. Results are presented in Figure \ref{fig1} (the Log of the $L^2$ norm is plotted). We observe a fast exponential decay, 
indicating that the two reservoirs synchronize 
by the time-step $k=20$.

\FloatBarrier

\subsection*{Time-series prediction.} 
\begin{figure}
\centerline{Trajectory 1}
\centerline{
\includegraphics[width=0.5\textwidth]{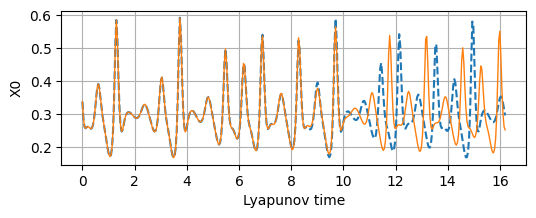}
\includegraphics[width=0.5\textwidth]{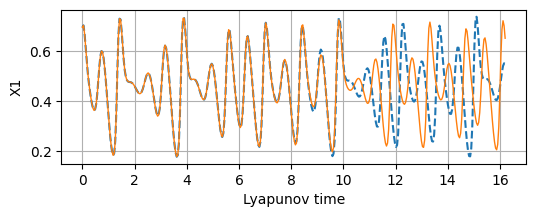}}
\centerline{
\includegraphics[width=0.5\textwidth]{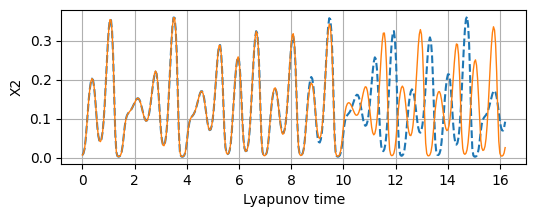}
\includegraphics[width=0.5\textwidth]{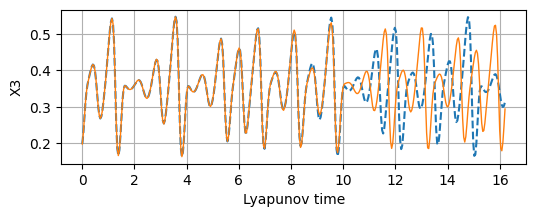}}
\centerline{Trajectory 2}
\centerline{
\includegraphics[width=0.5\textwidth]{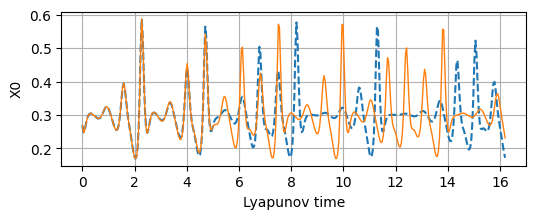}
\includegraphics[width=0.5\textwidth]{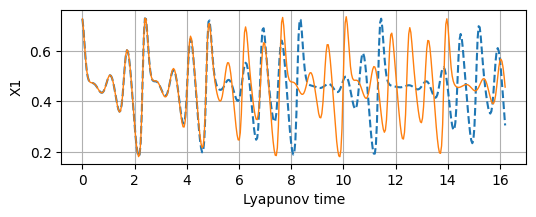}}
\centerline{
\includegraphics[width=0.5\textwidth]{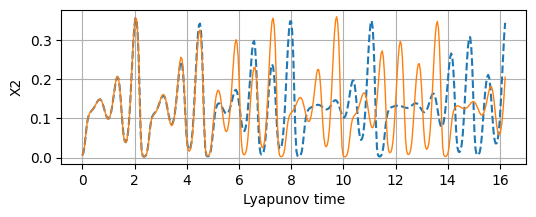}
\includegraphics[width=0.5\textwidth]{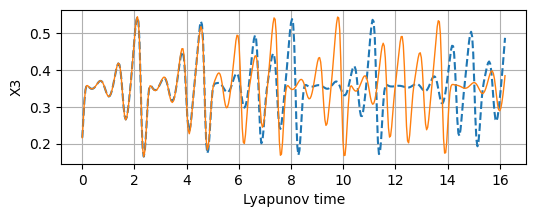}}
  \caption{Prediction of individual time-series with $\Delta t=2$. This demonstrates that the ESN predicts trajectories of the LV system for approximately $6,\ldots,8$ Lyapunov times. LV - solid yellow, ESN - dashed blue.}
  \label{fig2}
\end{figure}

An example of how the ESN predicts individual trajectories is depicted in Figure \ref{fig2}. In particular, Figure \ref{fig2} shows two different trajectories with different initial conditions, both lying on the attractor. 
We observe that the ESN can predict individual trajectories for a long time. In particular, 
The first trajectory is predicted for approximately 8 Lyapunov times, while the second trajectory is predicted for about 5 Lyapunov times. Overall, trajectories with initial conditions sampled from the attractor are predicted on average for about 7 Lyapunov times. When predicting individual trajectories of the LV system, we use a short warm-up period of 10 ESN time-steps. Thus, we generate a short "true" trajectory of the LV system on $t\in[0,20]$ and use this data in \eqref{res} to warm up reservoir states.
Moreover, the ESN generates trajectories with the time-step $\Delta t=2$, which yields a significant acceleration compared to the computational time-step of the LV model $\delta t=0.01$.
%

%

\FloatBarrier

\subsection*{Equilibrium Histogram prediction.}
Since the LV equations are a chaotic system, it is unrealistic to expect that the ESN would predict a chaotic LV  trajectory forever.  
This is confirmed in Figure \ref{fig2}.
\rev{However, we can ask three important questions regarding long-time simulations of the ESN: (i) whether the ESN remains stable, (ii) whether the ESN reproduces the bulk statistical properties of the attractor, and (iii) whether the ESN correctly captures the tails of the stationary distributions and the statistics of rare events.
The second question has been answered positively for several other chaotic systems 
(e.g. \cite{pathak2018model, antonik2018using,zhixin2018attractor,racca2021robust}), whereas the third question has received less attention.
}
\rev{One might consider accurate reproduction of the bulk statistical properties a necessary prerequisite for assessing the more demanding question of whether the ESN can reproduce the tails of the distribution and rare-event statistics.}

\begin{figure}[ht]
\centerline{
\includegraphics[width=0.9\textwidth]{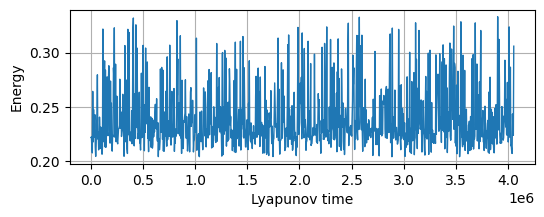}
}
\caption{Energy $E=\tfrac12 \sum x_i^2$ in long simulations of the ESN. The total simulation time is $T=4 \times 10^6$ Lyapunov times.}
\label{fig:ener}
\end{figure}
\begin{figure}
\centerline{
\includegraphics[width=0.52\textwidth]{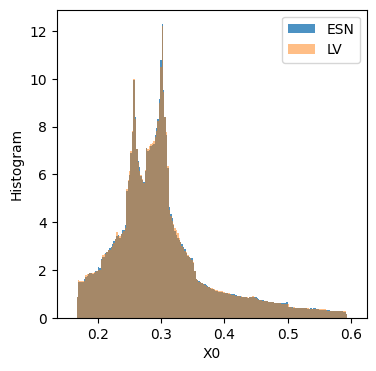}
\includegraphics[width=0.52\textwidth]{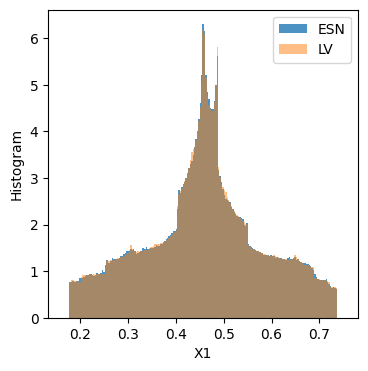}
  }
\centerline{
\includegraphics[width=0.52\textwidth]{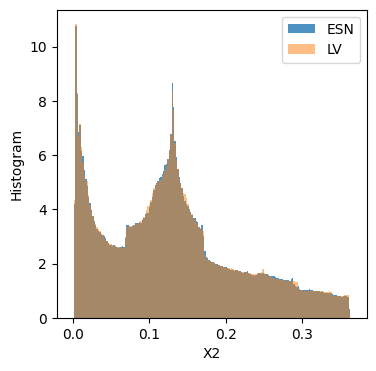}
\includegraphics[width=0.52\textwidth]{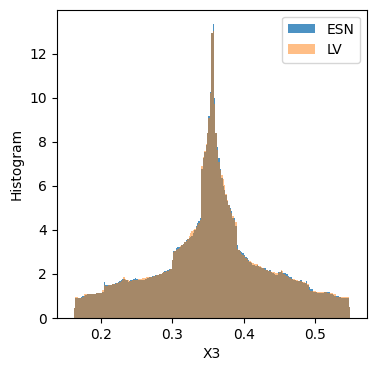}
}
\caption{Histograms of dependent variables $x_i$, $i=0,\ldots,3$ in long simulations of the LV model in \eqref{lv} and simulations of the ESN. 
\rev{The histograms generated by the LV system in \eqref{lv} and by the ESN agree very well, including in the tails of all four variables.}
}
\label{fig3}
\end{figure}

We verified numerically that the ESN does not blow up and does not collapse to a steady state in long-time simulations. \rev{In particular, Figure \ref{fig:ener} depicts energy $E=\tfrac12 \sum x_i^2$ in a long simulation of the ESN. The energy fluctuates in the interval $[0.2, 0.37]$ but the trajectory appears statistically stationary and does not collapse.}
Moreover, the ESN can generate trajectories that correctly represent the chaotic attractor of the LV model. Figure \ref{fig3} shows that the ESN correctly reproduces histograms for all four variables with a very high accuracy; the two histograms (LV and ESN) almost completely overlap everywhere.
All four histograms appear to have finite tails with well-pronounced cut-offs. Variable $x_0$ has a much longer, slowly-decaying right tail. Zoom-in on histograms' tails is presented in the supplementary material. 
\rev{Histograms in Figure \ref{fig3} are generated using $10^7$ sampled points with $\Delta t=2$.}

\FloatBarrier

\subsection*{Predictions for Block Maxima and Rare Events}
Next, we perform a quantitative comparison 
by computing the maximum over $q$ points in the trajectory, and fitting the Generalized Extreme Value (GEV) distribution. 
The GEV distribution is a three-parameter distribution with the probability density function
\[
f(x) = \frac{1}{\sigma} t(x)^{\xi+1} e^{-t(x)}, \quad
t(x) = \begin{cases}
    \left[ 1 + \xi \left( \frac{x-\mu}{\sigma}\right)\right]^{-1/\xi} & \xi\ne 0\\
    \exp\left( -\frac{x-\mu}{\sigma}\right) & 
    \xi = 0.
\end{cases}
\]
Here, the three parameters are
location $\mu$, scale $\sigma>0$, and shape $\xi$ with $1+\xi(x-\mu)/\sigma>0$. 
The shape parameter characterizes the tail of the distribution, and there are three distinct cases (Fisher–Tippett–Gnedenko theorem): (i) Fréchet distribution when $\xi > 0$
(heavy tails, e.g., Student’s t distribution with small degrees of freedom), (ii) Gumbel distribution when $\xi=0$ (light-tailed, e.g., exponential, normal), and (iii) Weibull distribution when $\xi < 0$ (bounded upper tail, e.g., uniform).  

Thus, after a long stationary trajectory is generated, the maximum is computed over non-overlapping windows, i.e.
\begin{equation}
\label{maxx}
y_i(l) = \max_{lq \le j < (l+1)q} x_i (j\Delta t), \qquad i=0,\ldots,3, \qquad 
l=0, \ldots, \Mm-1,
\end{equation}
and $y_i(l)$, $l=0,\ldots,\Mm-1$ are used for the GEV fit \rev{with $\Mm=\lfloor 10^7 / q \rfloor$.}
\rev{In this study, we do not define individual rare events using a fixed threshold. Instead, we examine upper-tail behavior through extreme-value statistics. For stationary trajectories, extremes are represented by the block maxima $y_i(l)$ in \eqref{maxx}. For transient ensemble simulations, they are represented by the finite-time maxima 
$z_k^{(i)}$ in \eqref{maxx2}.
We fit GEV distributions to these maxima to quantify and compare the upper-tail behavior of the LV system and the ESN.}

\rev{
In addition, we quantify how far the block maxima lie in the upper tails of the marginal stationary distributions. 
For each variable \(x_i\), block size \(q\), and \(p\in \{0.25,0.50,0.75\}\), we compute the $p$-th percentile \(\eta_i^{(p)}(q)\) of the observed block maxima 
together with the corresponding exceedance probability for $x_i$
estimated from its empirical stationary distribution. Specifically, denoting the block-maximum random variable by
\(m_i(q)\), we compute 
\(\eta_i^{(p)}(q)\) such that 
\begin{equation*}
\Pr\left(m_i(q) \le \eta_i^{(p)}(q) \right) = p,
\qquad i=0,\ldots,3.
\end{equation*}
We then compute the marginal exceedance probability
\begin{equation}
\label{eq:prob_eta}
P_i(q,p) = \Pr \left(x_i\geq\eta_i^{(p)}(q)\right),
\qquad i=0,\ldots,3
\end{equation}
using the empirical stationary distribution of $x_i$ generated by the LV simulations.  
The probabilities $P_i(q,p)$ quantify how frequently a randomly sampled state from the stationary distribution exceeds the corresponding percentile threshold of the block-maxima distribution.

The exceedance probabilities are presented in Table~\ref{tab:prob}.
All exceedance probabilities are very small. The small exceedance probabilities confirm that the percentile thresholds of the block maxima correspond to rare events in the stationary distributions. Across all variables, block sizes, and percentiles, the probabilities range from approximately $8.7\times10^{-4}$ to $5.1\times10^{-2}$. Thus, even the least extreme threshold is exceeded by only about $5\%$ of the stationary states, while the most extreme thresholds are exceeded by less than $0.1\%$. Moreover, $P_i(q,p)$ generally decreases as either $q$ or $p$ increases, because larger blocks and higher percentiles produce more extreme thresholds. These results demonstrate that our GEV analysis probes the right tails of the stationary distributions and therefore characterizes rare events.}

\begin{table}[]
\centering
\begin{tabular}{|l|c|c|c|}
\hline
& $p=0.25$, $q=50$ 
& $p=0.25$, $q=100$ 
& $p=0.25$, $q=200$  \\
\hline
$x_0$   
& 0.029 & 0.01 & 0.0052  \\
$x_1$  
& 0.032 & 0.012 & 0.0051  \\
$x_2$  
& 0.051 & 0.019 & 0.0073  \\
$x_3$    
& 0.05 & 0.018 & 0.0071  \\
\hline
\hline
& $p=0.5$, $q=50$ 
& $p=0.5$, $q=100$ 
& $p=0.5$, $q=200$  \\
\hline
$x_0$   
& 0.013 & 0.0053 & 0.003  \\
$x_1$  
& 0.014 & 0.0054 & 0.003  \\
$x_2$  
& 0.022 & 0.0079 & 0.0029  \\
$x_3$    
& 0.022 & 0.0076 & 0.0029  \\
\hline
\hline
& $p=0.75$, $q=50$ 
& $p=0.75$, $q=100$ 
& $p=0.75$, $q=200$  \\
\hline
$x_0$   
& 0.005 & 0.003 & 0.0013  \\
$x_1$  
& 0.005 & 0.003 & 0.0013  \\
$x_2$  
& 0.007 & 0.003 & 0.0013  \\
$x_3$    
& 0.005 & 0.003 & 0.0013  \\
\hline
\end{tabular}
\caption{Exceedance probabilities $P_i(q,p)$ in \eqref{eq:prob_eta}. Probabilities are computed from the LV data. Block maxima and corresponding probabilities computed from the ESN simulations follow a similar trend.}
\label{tab:prob}
\end{table}

The GEV fit can depend on the number of points in the block size, $q$. Recall that the sampling time-step is $\Delta t=2$. We present three cases with $q=50$, $100$, and $200$. This corresponds approximately to 2, 4, and 8 Lyapunov times.
Results of the GEV fit are presented in Table \ref{tab:gev}. 
\rev{Corresponding 95\% confidence intervals computed using parametric bootstrapping \cite{panagoulia2014stationary} 
are presented in Appendix \ref{sec:ci}.

In addition, we depict histograms for block maxima in Figure \ref{fig:histmax}.
The histograms of the block maxima obtained from the original data are reproduced very well by the ESN. In particular, the ESN captures both the location and spread of the distribution, the frequency of the largest observed values, and small left tails for $q=100$ and $q=200$. This close agreement indicates that the ESN reproduces not only the system’s typical behavior but also the statistical properties of its extreme events.}

\def\wfigt{0.32\textwidth}

\begin{figure}
\centerline{
    \includegraphics[width=\wfigt]{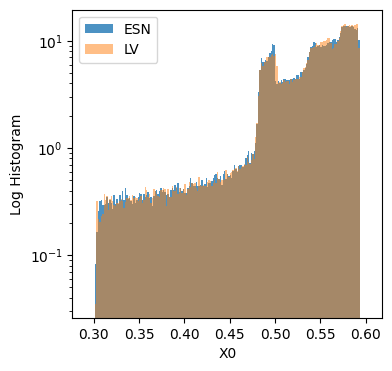}
    \includegraphics[width=\wfigt]{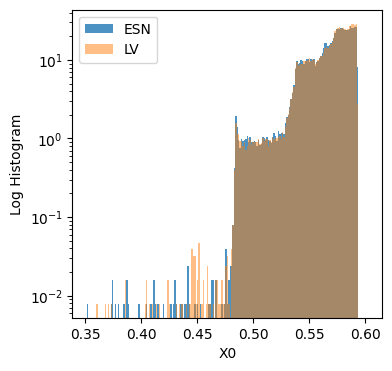}
    \includegraphics[width=\wfigt]{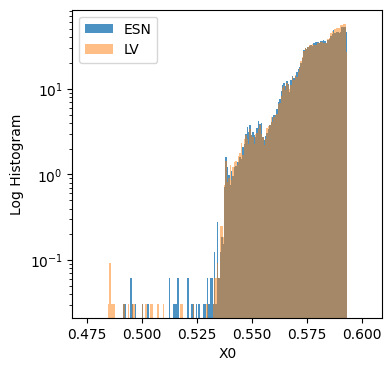}
  }
 \centerline{
    \includegraphics[width=\wfigt]{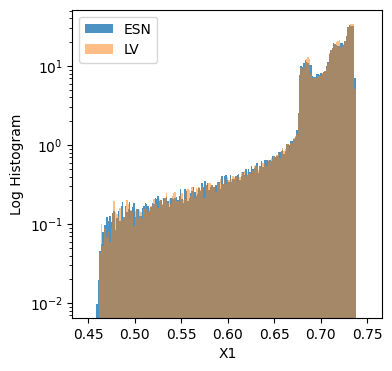}
    \includegraphics[width=\wfigt]{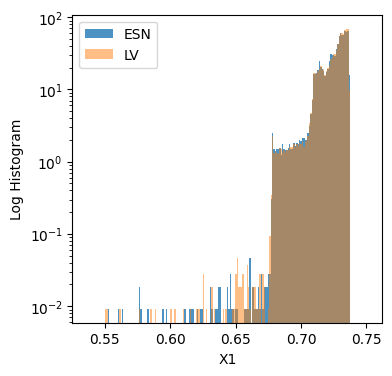}
    \includegraphics[width=\wfigt]{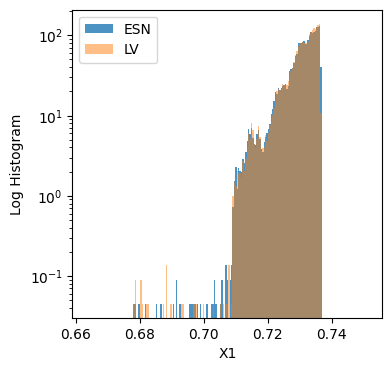}
  }
\centerline{
    \includegraphics[width=\wfigt]{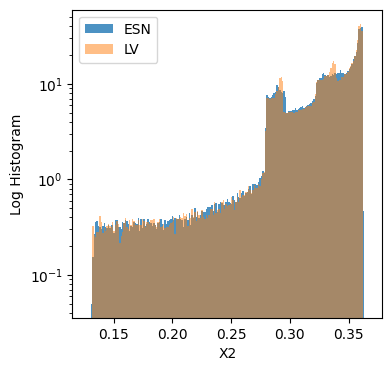}
    \includegraphics[width=\wfigt]{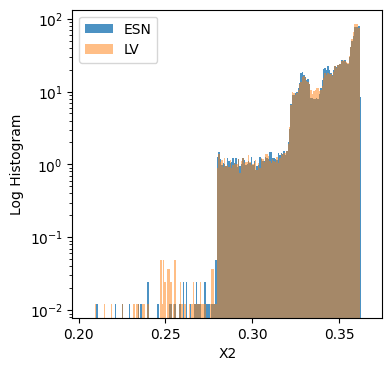}
    \includegraphics[width=\wfigt]{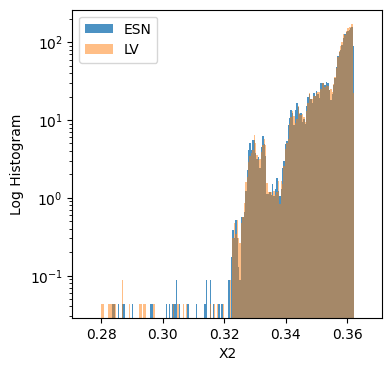}
  }
\centerline{
    \includegraphics[width=\wfigt]{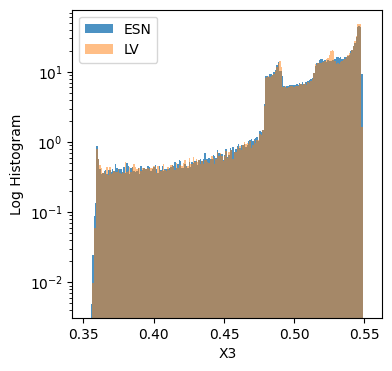}
    \includegraphics[width=\wfigt]{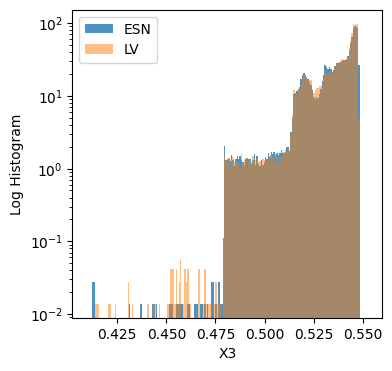}
    \includegraphics[width=\wfigt]{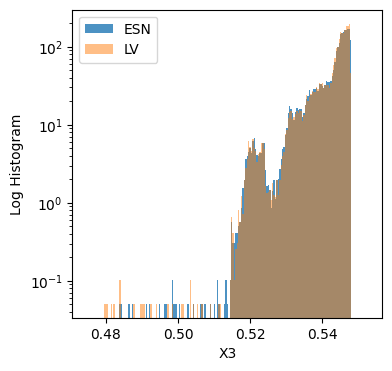}
  }
  \caption{Histograms of the block maxima $y_i$, $i=0,\ldots,3$ (top to bottom) for three values of the block size $q=50$ (left column), $q=100$ (middle column), and $q=200$ (right column).}
  \label{fig:histmax}
\end{figure}

\rev{Parameter estimation and confidence-interval construction for the GEV distribution with \(\xi<0\) are challenging because the distribution has a finite, parameter-dependent upper endpoint, \(\mu-\sigma/\xi\). Observations near this endpoint strongly constrain the admissible parameter values, resulting in an asymmetric and potentially irregular likelihood. These difficulties are particularly severe when \(\xi<-0.5\), for which the standard regularity conditions underlying maximum-likelihood inference fail, and the estimator may have a nonstandard asymptotic distribution; consequently, conventional standard errors and confidence intervals based on the Fisher information may be unreliable \cite{smith1985maximum}.
Additionally, it has been recognized that maximum-likelihood estimation can be numerically unstable for parameter-dependent-support models, with the likelihood potentially lacking a finite local maximum in some parameter regimes \cite{hosking1985estimation}
Although alternative estimators and bootstrap procedures have been developed \cite{hosking1985estimation,hosking1990lmoments,greenwood1979probability,gomes2016bootstrap,gilleland2020bootstrap1,gilleland2020bootstrap2}, inference remains sensitive to observations near the upper endpoint, particularly when the fitted shape parameter is near or below \(-1\), and the resulting estimates should therefore be interpreted with caution.}

\rev{In Tables~\ref{tab:gev} and~\ref{tab:gevci}, we report numerical GEV fits obtained using the \texttt{scipy.stats} Python package; this package uses notation $c = -\xi$. 
}
Results in Table \ref{tab:gev} indicate that the GEV fit is only weakly sensitive to the block length, $q$; \rev{However, Figure \ref{fig:histmax} indicates that there is a transition between $q=50$ and $q=100$. For $q \geq 100$, the histograms of the block maxima exhibit thin left tails. 
}
For all four variables, the GEV fit indicates finite tails, and there is a very good agreement between the simulations of the LV system and the trajectories generated by the ESN.

\rev{We use a parametric bootstrap procedure \cite{panagoulia2014stationary} to quantify the sampling variability of the estimated shape parameter, \(\xi\), and to construct nominal 95\% confidence intervals. These intervals summarize the variability of the complete estimation procedure under repeated samples from the fitted GEV model. Because inference is nonregular near \(\xi=-1\), however, their actual coverage may differ from the nominal 95\% level, and the intervals should therefore be interpreted with caution.}

\rev{
The empirical block-maxima distributions in Figure \ref{fig:histmax} provide direct evidence that the ESN reproduces the statistics of block maxima.
%
%
Therefore, we can conclude that the ESN accurately reproduces the upper tails and rare-event statistics of all four variables.
This agreement is particularly notable because reproducing tail-sensitive statistics provides a more demanding test of the ESN than reproducing bulk properties of the invariant measure.}

\begin{table}[]
\centering
\begin{tabular}{|l|l|l|l|l|l|l|l|l|l|}
\hline
   & \multicolumn{3}{|c|}{$q=50$} & \multicolumn{3}{|c|}{$q=100$} & \multicolumn{3}{|c|}{$q=200$} \\
   \hline
   & LV     & ESN    & Err    & LV     & ESN     & Err    & LV     & ESN     & Err    \\
   \hline
$x_0$ & 
-0.925 & -0.911 & 1.5\%  & 
-0.867 & -0.847 & 2.3\%  & 
-0.86 &  -0.834 & 3.0\% \\
$x_1$ & 
-1.077 & -1.079 & 0.18\%  & 
-0.937 & -0.923 & 1.5\%  & 
-0.901 &  -0.891 & 1.1\% \\
$x_2$ & 
-1.097 & -1.079 & 1.6\%  & 
-1.11 & -1.084 & 2.3\%  & 
-1.10  & -1.09 & 0.9\%  \\
$x_3$ & 
-1.086 & -1.064 & 2.0\%   
& -1.11 & -1.08 & 2.7\%  & 
-1.1 & -1.07 & 2.7\%  \\
\hline
\end{tabular}
\caption{Shape parameter $\xi$ for the GEV fit for right tails of histograms using $\max$ values computed with \eqref{maxx} and corresponding relative errors.}
\label{tab:gev}
\end{table}


\subsection*{Ensemble prediction.}
There is a large amount of literature on extreme events (e.g., \cite{lucarini2016extremes, coles2001introduction}, including in the weather and climate communities (see, for example, reviews \cite{ext2,ext6,ext5,ext7} and 
\cite{chattopadhyay2020data}).
In recent years, extreme events in models of competition have also attracted interest in the biological community (e.g. \cite{ext1,ext2,ext3,ext4}).
\rev{These studies motivate our analysis of extreme-event statistics, although the numerical experiments presented here are restricted to the low-dimensional LV system.
In this section, we investigate how well the ESN reproduces extreme-event statistics in ensemble simulations of the LV model.}
\rev{Unlike the stationary analysis, this experiment tests whether the ESN can reproduce finite-time extreme statistics in non-equilibrium ensemble simulations of the LV system.}
To this end, we select a point $m \in \R^4$ on the attractor, and consider an ensemble of $MC = 5,000$ initial conditions generated as $\bx^{(i)} \in Unif[m-v, m+v]$, where $v=C \rho \in \R^4$ and $\rho_k$ is the range of the variable $x_k$ on the attractor (same as the range of the equilibrium distribution in Figure \ref{fig3}). Next, we compute 
\begin{equation}
\label{maxx2}
z_k^{(i)} = \max_{0 \le t \le T}(x_k(t) \text{~with~} \bx(0) = \bx^{(i)}), \qquad 
k = 0, \ldots, 3, \qquad
i = 1,\ldots,MC.
\end{equation}
Here we present one particular case with 
$m=[0.30, 0.45, 0.13, 0.36]$, $\rho \approx [0.43, 0.53, 0.36, 0.39]$, $T=300$, and $C=0.3$. The scaling of the width with $\rho$ was chosen to relate the variability of the ensemble with the variability in stationary simulations, but the magnitude of $\rho$ is approximately the same for all four variables. 
We also determined numerically that the appropriate range for the constant is $0 < C \lesssim 0.4$. However, for small values of $C \le 0.2$ the distribution of $z_k^{(i)}$
becomes too narrow for an accurate GEV fit. Nevertheless, for tight ensembles with $C \le 0.2$, ESN ensemble simulations reproduce histograms of $z_k^{(i)}$ well.
With $C> 0.4$, the ESN's performance deteriorates, since it has not been trained to reproduce transient behavior away from the attractor.

Figure \ref{fig4} depicts histograms of $z_k^{(i)}$ 
and the corresponding values of the shape parameter from the GEV fit are presented in Table \ref{tab:gev2}. Figure \ref{fig4} and Table \ref{tab:gev2} demonstrate that
ESN quantitatively reproduces the statistics of extremes in ensemble simulations. We verified that this behavior is generic for 10 different points $m$
(centers of the initial distribution). Overall, the ESN reproduces the statistics of maxima with approximately 10\%–20\% accuracy, with overlapping confidence intervals.
As we discussed before, the GEV fit is not reliable for narrow ensembles $C<0.2$, but visually, the ESN reproduces histograms of block maxima in this regime very well.

\rev{These results show that the ESN accurately reproduces transient extreme-event statistics when the ensemble remains close to the underlying attractor. However,
the ESN's performance deteriorates as the ensemble spread increases and trajectories explore regions of phase space that are not represented in the training data.}

\def\wfig4{0.35\textwidth}

\begin{figure}
\centerline{
    \includegraphics[width=\wfig4]{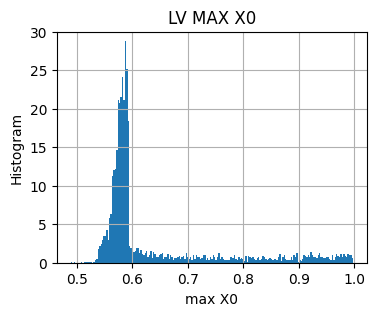}
    \includegraphics[width=\wfig4]{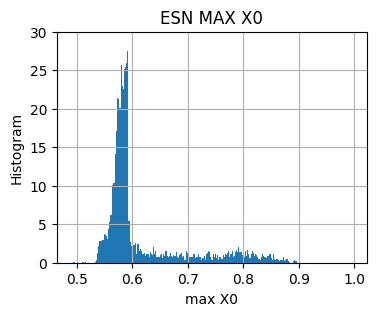}
  }
  \centerline{
    \includegraphics[width=\wfig4]{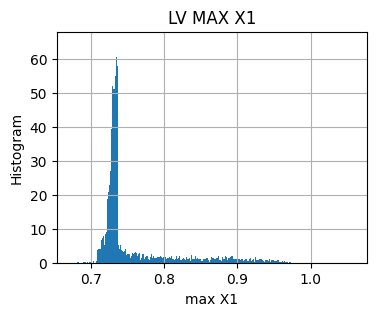}
    \includegraphics[width=\wfig4]{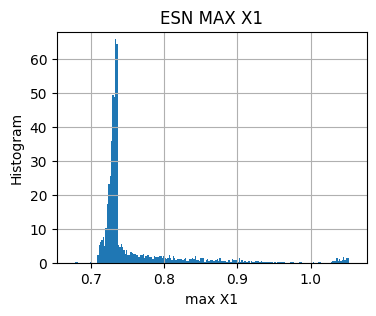}
}
\centerline{
    \includegraphics[width=\wfig4]{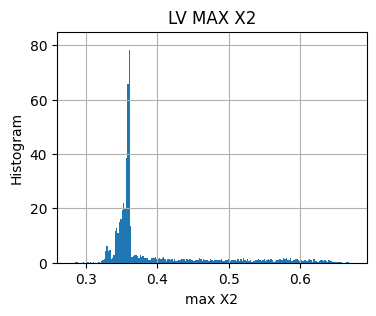}
    \includegraphics[width=\wfig4]{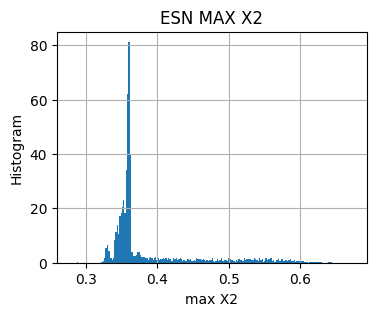}
  }
  \centerline{
    \includegraphics[width=\wfig4]{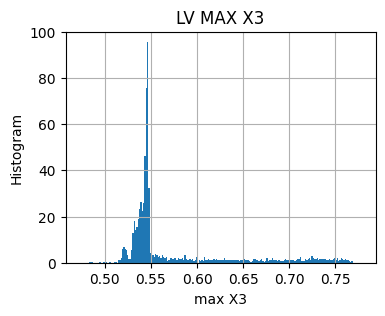}
    \includegraphics[width=\wfig4]{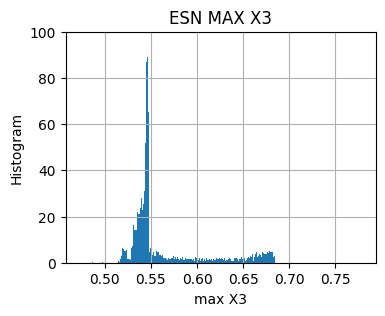}
}
  \caption{Comparison of histograms of 
  $z_k^{(i)} = \max_{0 \le t \le T}(x_k(t) \text{~with~} \bx(0) = \bx^{(i)})$, $i = 1,\ldots,MC$ in ensemble simulations with $T=300$ and $\Delta t=2$ of the LV and ESN models with initial conditions centered around  $m=[0.30, 0.45, 0.13, 0.36]$. The corresponding GEV fits for the shape parameter $\xi$ are presented in Table \ref{tab:gev2}.}
  \label{fig4}
\end{figure}

\begin{table}[]
\centering
\begin{tabular}{|l|c|c|}
\hline
   &  LV     & ESN    \\
   \hline
$z_0$ & $0.35 \pm 0.066$ & $0.28 \pm 0.088$ \\
$z_1$ & $0.30 \pm 0.047$ & $0.26 \pm 0.123$  \\
$z_2$ & $0.30 \pm 0.039$ & $0.27 \pm 0.122$   \\
$z_3$ & $0.29 \pm 0.108$ & $0.24 \pm 0.284$   \\
\hline
\end{tabular}
\caption{Shape parameter $\xi$ for the GEV fit for the distribution of 
$z_k = \max_{0 \le t \le T}(x_k(t))$ in ensemble simulations 
with $T=300$ and $\Delta t=2$ of the LV and ESN models 
with initial conditions centered around  $m=[0.30, 0.45, 0.13, 0.36]$.
Corresponding histograms are depicted in Figure \ref{fig4}.
\rev{The uncertainties represent 95\% confidence intervals computed by standard non-parametric bootstrapping and normal approximation \cite{tibshirani1993introduction}.}}
\label{tab:gev2}
\end{table}

\FloatBarrier

\subsection*{Predictions under Incomplete Observations}
\begin{figure}
\centerline{Trajectory 1}
\centerline{
\includegraphics[width=0.5\textwidth]{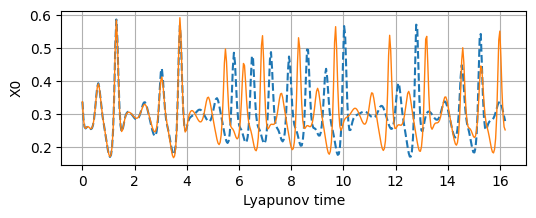}
\includegraphics[width=0.5\textwidth]{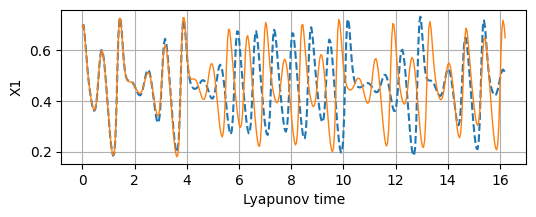}}
\centerline{
\includegraphics[width=0.5\textwidth]{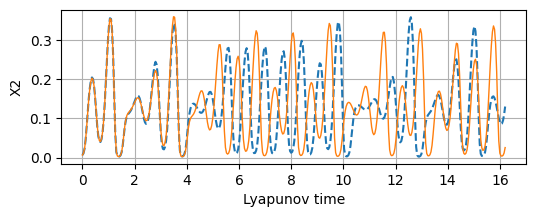}
\includegraphics[width=0.5\textwidth]{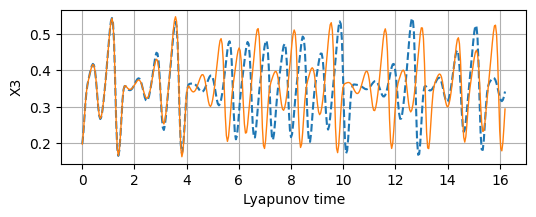}}
\centerline{Trajectory 2}
\centerline{
\includegraphics[width=0.5\textwidth]{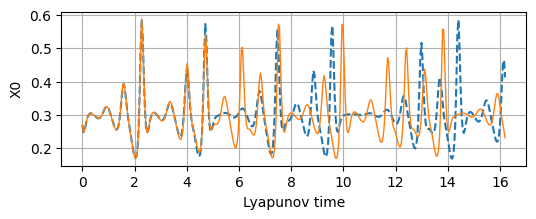}
\includegraphics[width=0.5\textwidth]{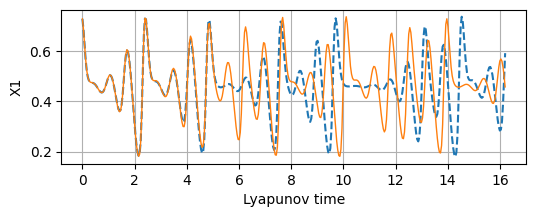}}
\centerline{
\includegraphics[width=0.5\textwidth]{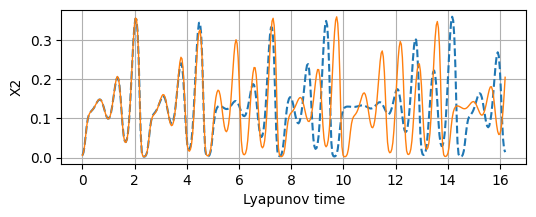}
\includegraphics[width=0.5\textwidth]{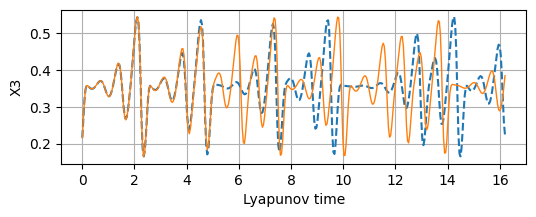}}
  \caption{Prediction of individual time-series with $\Delta t=2$ under incomplete observations with $\xt=\{x_0, x_1, x_2\}$. Compare with Figure \ref{fig2} for predictions with complete observations. LV - solid yellow, ESN - dashed blue.}
  \label{fig:trajinc}
\end{figure}

\rev{We also analyze whether the ESN is capable of predicting the behavior of dynamical variables under incomplete observations. To this end, we use a smaller input vector in \eqref{res}, $\xt \in \R^\Nt$ with $\Nt < N$, but predict all dynamical variables, $x \in \R^N$. There are many possible combinations for selecting a subset of dynamical variables $\xt$. Here we present one particular case where we use a three-dimensional input vector $\xt=\{x_0,x_1,x_2\}$. 
We depict the prediction of individual trajectories with incomplete observations in Figure \ref{fig:trajinc}; here we use the same initial conditions as in Figure \ref{fig2} for comparison.
The prediction deteriorates significantly, which is particularly evident for Trajectory 1, which is predicted accurately only for approximately 4 Lyapunov times under incomplete observations. Surprisingly, the prediction time for Trajectory 2 does not deteriorate significantly in this case.

We also tested other configurations for the input vector $\xt$. Similar behavior is observed for other combinations with $\xt \in \R^3$. Thus, 
we expect the ESN to predict trajectories for approximately 4 Lyapunov times under partial observations with $\xt \in \R^3$.  For $\xt \in \R^2$, the prediction times deteriorate further, and the ESN predicts individual trajectories for approximately one Lyapunov time.}

\begin{figure}
\centerline{
\includegraphics[width=0.52\textwidth]{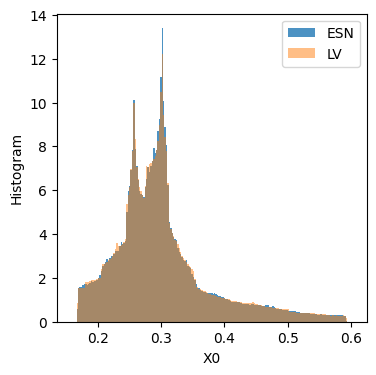}
\includegraphics[width=0.52\textwidth]{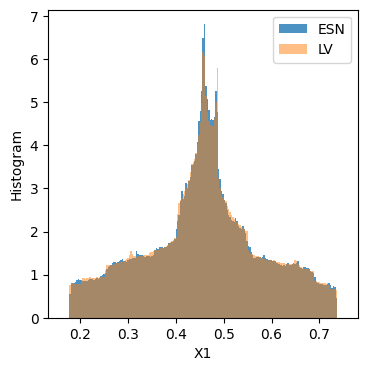}
  }
\centerline{
\includegraphics[width=0.52\textwidth]{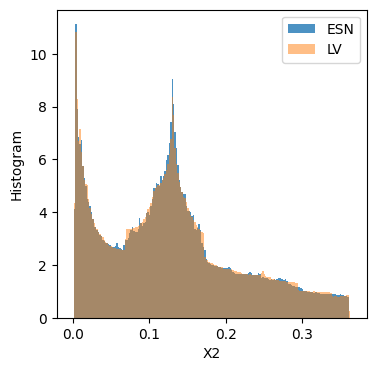}
\includegraphics[width=0.52\textwidth]{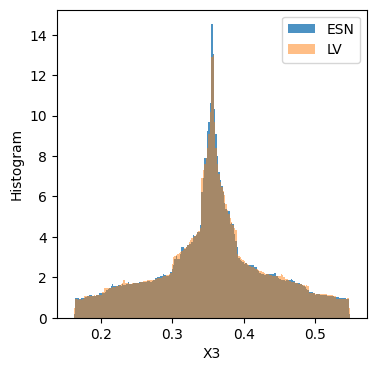}
}
\caption{Histograms of dependent variables $x_i$, $i=0,\ldots,3$ in long simulations of the LV model in \eqref{lv} and simulations of the ESN model under incomplete observations with $\xt=\{x_0, x_1, x_2\}$. Compare with Figure \ref{fig3} for predictions under complete observations. The $L^2$ errors are reported in Table \ref{tab:ngrc}.
}
\label{fig:histinc}
\end{figure}

\rev{Furthermore, we depict the histograms of the dynamical variables in Figure~\ref{fig:histinc}. This figure demonstrates that the ESN is capable of reproducing these histograms with high accuracy under incomplete observations with $\xt \in \R^3$. We also verified that this behavior is generic for $\xt \in \R^3$. For $\xt \in \R^2$, histogram predictions deteriorate further, but the ESN retains some prediction power. In particular, the ESN reproduces the overall shape of the distribution (long tails, position of peaks, etc.), but discrepancies become larger.

Overall, our results demonstrate that the ESN retains some predictive capability under incomplete observations. Although its ability to predict individual trajectories deteriorates significantly, it continues to reproduce the marginal distributions of the dynamical variables with relatively high accuracy. As expected, both trajectory prediction and statistical reconstruction deteriorate as the dimension of the observed input decreases.}

\FloatBarrier

\subsection*{Comparison with Next Generation Reservoir Computing.}
\begin{figure}
\centerline{
\includegraphics[width=0.52\textwidth]{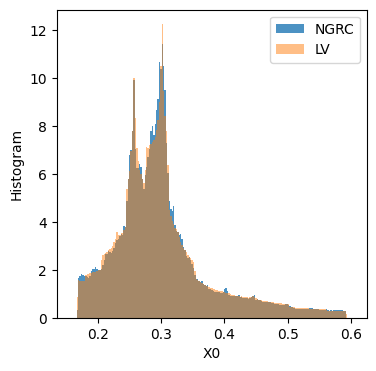}
\includegraphics[width=0.52\textwidth]{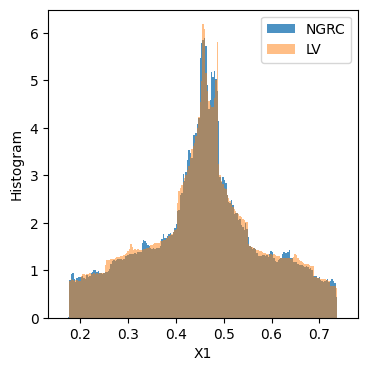}
  }
\centerline{
\includegraphics[width=0.52\textwidth]{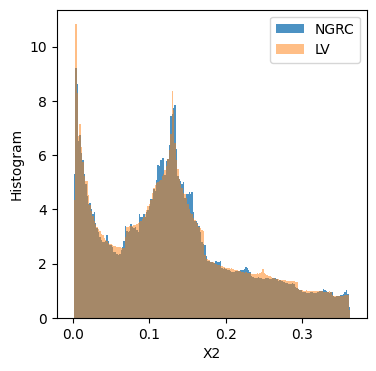}
\includegraphics[width=0.52\textwidth]{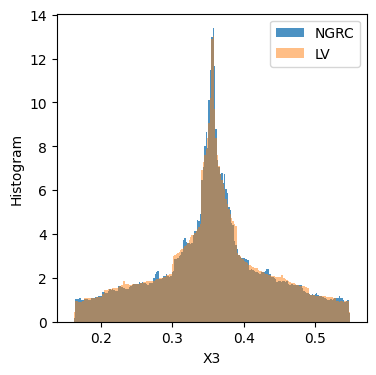}
}
\caption{Histograms of dependent variables $x_i$, $i=0,\ldots,3$ in long simulations of the LV model in \eqref{lv} and simulations of the NGRC model with $p=4$ and $k=2$. Comparison with Figure \ref{fig3} demonstrates that the performance of the ESN is superior to the NGRC performance depicted here.
}
\label{fig5}
\end{figure}

\rev{
An NGRC model \cite{gauthier2021ngrc}
is defined as a linear map 
$\bx_{i+1} = \bx_i + W_{out} \, \bbO_i$, where $W_{out}$ is a trainable output matrix
computed similarly to \eqref{W_out formula}, $\bx_i \equiv \bx(i\Delta t) \in \R^N$ are the dynamical variables of interest (i.e., predictions for the LV model variables in our case), reservoir 
$\bbO_i = [c \, \oplus \, \bbO_{lin, i} \, \oplus \, \bbO_{nlin, i}^{(p)}]$ that is defined as an extended state vector of dynamical variables, and $c$ is a constant; we use $c=1$. Here $\oplus$ denotes the vector concatenation operation.
NGRC models rely on including memory terms into the state vector. Define $\br_i = \bx_i \oplus \bx_{i-1} \oplus\ldots \oplus \bx_{i-(k-1)}$ that includes the current 
state and $k-1$ previous states. 
The linear part of the extended state vector is  $\bbO_{lin, i} = \br_i$, and the nonlinear part includes 
monomials of degree $p$, allowing repeated factors
but excluding duplicate permutations (including each distinct monomial only once), i.e.,
$\bbO_{nlin, i}^{(p)} = [\br_i \otimes \br_i \otimes \ldots \otimes \br_i]$ with $\br_i$ appearing $p$ times, where $\otimes$ denotes the outer product.
Here, $k$ and $p$ are two design parameters of the NGRC model.
We refer readers to \cite{gauthier2021ngrc} for details.

We compared the performance of the ESN model 
in \eqref{ESN} with three different NGRC models. We selected parameters $k$ and $p$ so that 
the dimensions of the corresponding NGRC extended state vectors, $\bbO_i$, are comparable to the ESN reservoir size, $D=200$
Table \ref{tab:ngrc} presents a comparison of $L^2$ errors  
between the marginal stationary histograms generated by the LV system and those generated by the ESN and NGRC models, as well as the approximate trajectory-prediction times.
In addition, Figure~\ref{fig5}
compares the marginal histograms generated by the NGRC model with $p=4$ and $k=2$ with those generated by the LV simulations. Individual trajectory predictions are not shown for brevity.
For the configurations considered here, the ESN outperforms all three NGRC models in reproducing the marginal stationary distributions and predicting individual trajectories. Among the NGRC models, the configuration with $p=4$ and $k=2$ provides the best overall performance.
}
\begin{table}[ht]
\centering
\begin{tabular}{|l|c|c|c|c|c|c|}
\hline
Model & Res. Size & Hist $x_0$ & Hist $x_1$ & Hist $x_2$ & Hist $x_3$ & Pred. Time\\
\hline
NGRC $p=3$, $k=2$ & 129 & 5.4  & 5.75 & 8.1 & 7.86 & $2$--$3$  \\
NGRC $p=3$, $k=3$ & 377 & 10.5 & 5.0 & 8.5 & 7.8 & $2$--$3$  \\
NGRC $p=4$, $k=2$ & 339 & 4.0  & 2.7 & 3.9 & 3.9 & $3$--$4$  \\
\hline
ESN with $\xt=\{x_0, x_1, x_2\}$ & 200 & 2.2 & 1.8 & 2.4 & 2.85  & $4$--$5$ \\
\hline
ESN                    & 200 & 0.87  & 0.63 & 1.03 & 1.01 & $6$--$10$ \\
\hline
\end{tabular}
\caption{Comparison of three NGRC models, ESN with incomplete observations, and the ESN. Reservoir size for each model,
$L^2$ errors in histograms of each variable, and approximate Lyapunov times for predicting trajectories. All histograms are computed with 200 bins.}
\label{tab:ngrc}
\end{table}

\FloatBarrier

\section{Conclusions}
We use a standard Echo-State Network to reproduce the behavior of the four-dimensional competitive Lotka--Volterra (LV) system in the chaotic regime. \rev{The main emphasis here is on investigating whether an ESN can reproduce distributional tails and rare-event statistics. The marginal stationary distributions of the LV system have bounded but slowly decaying tails, unlike the more rapidly decaying tails of the Lorenz--63 and Kuramoto--Sivashinsky examples. 
Therefore, the LV model provides an interesting test case to analyze how well ESN can learn tails of stationary distributions in chaotic systems and reproduce rare events.}

\rev{As expected from previous studies, the ESN predicts individual trajectories for several Lyapunov times and accurately reproduces the bulk statistical properties of the LV attractor. More importantly, the GEV fits show that the ESN reproduces the distribution of block maxima with high accuracy. 
We also show that the small exceedance probabilities of the block-maxima thresholds confirm that these maxima correspond to rare events in the stationary distributions.
These results demonstrate that an ESN can successfully extend its predictions from bulk attractor properties to rare-event statistics. Our results provide supporting evidence for the use of ESNs in low-dimensional chaotic systems that require the generation of trajectories with prescribed tail-sensitive statistical properties.}

We also compare transient ensemble simulations of the LV and ESN models. \rev{The ESN accurately reproduces the distributions of finite-time extremes when the ensemble remains close to the underlying attractor. However, the ESN's performance deteriorates as the initial ensemble becomes more spread out and trajectories explore regions of phase space that are atypical for long stationary simulations.}

\rev{Under incomplete observations, the ESN retains a meaningful but reduced predictive capability. With three observed variables, individual trajectories are typically predicted accurately for approximately four Lyapunov times, whereas reducing the input to two variables shortens this horizon to approximately one Lyapunov time. Statistical predictions are more robust: the ESN reproduces the main features of the marginal distributions - including their peaks and long tails - even when some variables are unobserved, although discrepancies increase as the input dimension decreases. These results demonstrate that the ESN can recover both dynamical and statistical information from partial observations, with the statistical features being more robust.}

\rev{We also compared the ESN with three NGRC configurations having extended state-vector dimensions comparable to the ESN reservoir size. For the configurations considered, the ESN consistently outperformed the NGRC models in both individual trajectory prediction and reconstruction of the marginal stationary distributions. Among the NGRC models, the configuration with $p=4$ and $k=2$ achieved the best overall performance. These results indicate that, for the LV system and parameter settings examined here, the ESN provides a more accurate representation of both the short-term dynamics and long-term statistical behavior.}

\rev{Taken together, our results demonstrate that an ESN trained on time-series data can reproduce not only bulk attractor statistics but also tail-sensitive quantities, including block-maxima distributions and finite-time extremes. This capability is strongest when the simulated trajectories remain within regions of phase space represented in the training data. The findings motivate further investigation of whether comparable statistical fidelity can be achieved in higher-dimensional and more complex dynamical systems.}

\smallskip

\noindent
Numerical code and data that support the findings of this study are openly available in \cite{code}.

\subsection*{Author Contribution}
A.E.: code development, simulations, visualization.
B.N.: conceptualization, manuscript writing.
I.T.: conceptualization, initial code development, student supervision, manuscript writing, final draft.

\section*{Acknowledgments}
A.E. acknowledges support from the 
UH Provost's Undergraduate Research Scholarship (PURS) and the UH
Summer Undergraduate Research Fellowship (SURF).
The authors thank Dr. M. Nicol for helpful discussions about the generalized extreme value distribution.

\appendix
\section{Synchronization Algorithm}
\label{sec:alg}
Here we present a short description of the synchronization test used to test independence of the ESN from initial conditions.
\begin{algorithm}[h]
\caption{Testing Independence of the Reservoir on Initial Conditions}
\label{alg1}
\begin{algorithmic}
\item
Generate $W_{in}$ and A
\item
Reservoir 1: $r_1(0) = 0$
\item
Reservoir 2: $r_2(0) = \epsilon \times N(0,1)$
\item 
M=50
\item 
L2norm[0] = $||r_1(0) - r_2(0)||_2$ 
\For{\texttt{k=1:M}}
\item 
\hspace{0.5cm} $r_1(t+\Delta t) = f(Ar_1(t) + W_{in} x(t))$ 
\item 
\hspace{0.5cm} $r_2(t+\Delta t) = f(Ar_2(t) + W_{in} x(t))$
\item 
\hspace{0.5cm} L2norm[k] = $||r_1(t+\Delta t) - r_2(t+\Delta t)||_2$
\EndFor 
\end{algorithmic}
\end{algorithm}

\FloatBarrier

\section{Confidence Intervals for Estimation of the Shape Parameter}
\label{sec:ci}

Here we present confidence intervals for the shape parameter $\xi$ of the GEV distribution, computed using parametric bootstrapping \cite{panagoulia2014stationary}. First, the GEV parameters are estimated from the standardized LV or ESN block-maxima data, with zero mean and unit variance; the shape parameter is invariant under the transformation $(Y-\bar{Y})/\text{std}(Y)$. Next, 2,000 synthetic datasets, each having the same size as the original dataset, are sampled from the fitted GEV distribution, and the shape parameter is re-estimated for each dataset. The 95\% confidence interval is then obtained from the 2.5th and 97.5th percentiles of the resulting 2,000 shape-parameter estimates.

We would like to point out that parameter estimation and the computation of confidence intervals for the GEV distribution with shape parameter $\xi\approx-1$ are delicate tasks
(see, e.g., \cite{gomes2016bootstrap,smith1985maximum,gilleland2020bootstrap2}). In this regime, the distribution has a finite, parameter-dependent upper endpoint, and the usual regularity conditions underlying maximum-likelihood inference are not fully satisfied. Consequently, the bootstrap intervals may not achieve their nominal coverage exactly and should therefore be interpreted with some caution. Nevertheless, they provide a useful indication of the uncertainty and the consistency of the observed trends.
\begin{table}[h]
\centering
\begin{tabular}{|c|c|c|c|c|c|c|}
\hline
   & \multicolumn{2}{|c|}{$q=50$} & \multicolumn{2}{|c|}{$q=100$} & \multicolumn{2}{|c|}{$q=200$} \\
   \hline
   & LV     & ESN    & LV     & ESN      & LV     & ESN     \\
   \hline
$x_0$ & 
[-0.93, -0.921] & [-0.915, -0.908] &  
[-0.871, -0.863] & [-0.851, -0.843]   & 
[-0.866, -0.854] &  [-0.84, -0.829]  \\
$x_1$ & [-1.091, -1.062] & [-1.097, -1.063] &  
[-0.944, -0.932] & [-0.93, -0.918] &  
[-0.907, -0.894] &  [-0.898, -0.885]  \\
$x_2$ & [-1.105, -1.083] & [-1.087, -1.056] & 
[-1.14, -1.093] & [-1.091, -1.064] & 
[-1.111, -1.086]  & [-1.105, -1.071]   \\
$x_3$ & [-1.095, -1.081] & [-1.077, -1.051] & 
[-1.111, -1.087] & [-1.079, -1.05] & 
[-1.105, -1.076] & [-1.097, -1.059]   \\
\hline
\end{tabular}
\caption{95\% Confidence Intervals for estimation of the shape parameter $\xi$ for the GEV fit for right tails of histograms using $\max$ values computed with \eqref{maxx}.}
\label{tab:gevci}
\end{table}

\newpage


\end{document}